# EXPLORING MAGNETIC ANISOTROPY IN BCC-STRUCTURED FERROMAGNETIC THIN FILMS WITH THREE SPIN LAYERS USING THE FOURTH ORDER PERTURBED HEISENBERG HAMILTONIAN

M.S.M. Farhan

Department of Physical Science, Trincomalee Campus, Eastern University, Sri Lanka.
Postgraduate Institute of Science, University of Peradeniya, Peradeniya, Sri Lanka.

**Abstract:**

This study investigates into the analysis of ferromagnetic thin films with a body centered cubic lattice and three spin layers, utilizing the solution of the fourth-order perturbed Heisenberg Hamiltonian equation with seven magnetic energy parameters. Spin-exchange interaction, magnetic dipole interaction, second-order magnetic anisotropy, fourth-order magnetic anisotropy, applied magnetic field, demagnetization energy, and stress-induced anisotropy were all taken into account. According to 3D plots, the minimum order of energy was observed when the second order magnetic anisotropy constant in the middle spin layer is less than those of the bottom and top spin layers. In all cases, the values of stress-induced anisotropy at the maxima of 3D plots are exactly the same when the values of the second-order magnetic anisotropy constants of the bottom, middle, and top spin layers are interchanged with each other. In 2D plots, the angle between consecutive magnetic easy and hard directions is approximately 90 degrees in all cases. Additionally, the magnetic easy and hard directions were observed to have exactly the same values when the second-order magnetic anisotropy constant of the spin layers' changes. These results were compared with the results obtained using the second and third order perturbed Heisenberg Hamiltonian.

## 1. Introduction:

Ferromagnetic thin films have emerged as crucial components in a variety of technological applications owing to their unique magnetic properties. These materials exhibit spontaneous magnetization and retain a strong magnetic moment even in the absence of an external magnetic field, making them highly desirable for applications such as data storage, magnetic sensors, and spintronics. Understanding and manipulating the magnetic properties of ferromagnetic thin films are essential for optimizing their performance in various devices.

The behavior of ferromagnetic thin films is governed by phenomena such as magnetic anisotropy, which influences the preferred direction of magnetization within the film. Anisotropy arises from various sources, including crystal lattice structure, interfacial effects, and external influences like stress or magnetic fields. Manipulating these anisotropies becomes a key aspect in tailoring the magnetic properties of thin films for specific applications.

The thickness of ferromagnetic thin films plays a crucial role in determining their magnetic behavior. As the film thickness decreases, quantum size effects and interface interactions become more pronounced, leading to modifications in magnetic properties. Researchers explore these nanoscale effects to design films with enhanced performance, such as increased coercivity, improved stability, and reduced power consumption.

The fabrication techniques employed in depositing thin films, such as sputtering, evaporation, or chemical vapor deposition, contribute to the microstructure and, consequently, the magnetic properties of the films. The film's composition, grain structure, and defects influence magnetic behavior, and researchers continually strive to optimize deposition methods to achieve desired magnetic characteristics.

In recent years, advances in theoretical models and computational simulations have provided valuable insights into the intricate details of the magnetic properties of ferromagnetic thin films. The interplay of factors such as exchange interactions, anisotropy constants, and magnetoelastic effects is explored to predict and understand the observed behaviors, guiding the design of novel materials for specific applications.

Ferromagnetic films have been described using various models. The quasistatic magnetic hysteresis of ferromagnetic thin films grown on a vicinal substrate has been theoretically investigated using Monte Carlo simulations [1]. Magnetic properties of ferromagnetic thin films with alternating superlayers were studied using the Ising model [2]. Structural and magnetic properties of two-dimensional FeCo ordered alloys have been determined by first-principles band structure theory [3]. Magnetic layers of Ni on Cu have been theoretically studied using the Korringa-Kohn-Rostoker Green's function method [4].

The experimental values of magnetic moments of ferromagnetic materials differ from the theoretical predictions due to the overlapping sub-shells. Magnetic thin films are heated and subsequently cooled during the annealing process. Throughout the heating and cooling process, stress-induced magnetic anisotropy (Ks) occurs due to the difference between the thermal expansion coefficients in the magnetic film and substrate. For soft magnetic materials, stress-induced anisotropy is comparable to the magneto-crystalline anisotropy. Therefore, coercivity depends on stress-induced anisotropy [5]. Research papers on the magnetic properties of ferromagnetic and ferrite thin and thick films have been previously published by some researchers [6-13].

In this manuscript, the fourth-order perturbed Heisenberg Hamiltonian with all seven magnetic energy parameters was solved for body-centered cubic structured ferromagnetic films with three spin layers. This represents the first research endeavor related to the effect of fourth-order perturbation on ferromagnetic films with three spin layers and a bcc structure. When the films are highly non-oriented, the influence of fourth or higher order perturbations has to be taken into account. Stress-induced anisotropy plays a crucial role in explaining the magnetic properties of materials with low crystal anisotropy [5]. A MATLAB computer was used to plot 3D and 2D graphs of energy versus stress-induced anisotropy and azimuthal angle of spin.

## 2. Model:

The Heisenberg Hamiltonian of ferromagnetic films can be formulated as following [8-10].

$$H = -\frac{J}{2}\sum_{m,n} \vec{S}_m \cdot \vec{S}_n + \frac{\omega}{2}\sum_{m \neq n}\left(\frac{\vec{S}_m \cdot \vec{S}_n}{r_{mn}^3} - \frac{3(\vec{S}_m \cdot \vec{r}_{mn})(\vec{r}_{mn} \cdot \vec{S}_n)}{r_{mn}^5}\right) - \sum_m D_{\lambda_m}^{(2)}(S_m^z)^2 - \sum_m D_{\lambda_m}^{(4)}(S_m^z)^4$$
$$- \sum_{m,n}[\vec{H} - (N_d \vec{S}_n/\mu_0)] \cdot \vec{S}_m - \sum_m K_s \sin 2\theta_m$$

Here $\vec{S}_m$ and $\vec{S}_n$ are two spins. Above equation can be simplified into following form

$$E(\theta) = -\frac{1}{2}\sum_{m,n=1}^{N}\left[\left(JZ_{|m-n|} - \frac{\omega}{4}\Phi_{|m-n|}\right)\cos(\theta_m - \theta_n) - \frac{3\omega}{4}\Phi_{|m-n|}\cos(\theta_m + \theta_n)\right]$$
$$- \sum_{m=1}^{N}(D_m^{(2)}\cos^2\theta_m + D_m^{(4)}\cos^4\theta_m + H_{in}\sin\theta_m + H_{out}\cos\theta_m)$$
$$+ \sum_{m,n=1}^{N}\frac{N_d}{\mu_0}\cos(\theta_m - \theta_n) - K_s \sum_{m=1}^{N}\sin 2\theta_m \qquad (1)$$

Here $N$, $m$ (or $n$), $J$, $Z_{|m-n|}$, $\omega$, $\Phi_{|m-n|}$, $\theta_m(\theta_n)$, $D_m^{(2)}$, $D_m^{(4)}$, $H_{in}$, $H_{out}$, $N_d$ and $K_s$ are total number of layers, layer index, spin exchange interaction, number of nearest spin neighbors, strength of long range dipole interaction, partial summations of dipole interaction, azimuthal angles of spins, second and fourth order anisotropy constants, in plane and out of plane applied magnetic fields, demagnetization factor and stress induced anisotropy constants respectively.

The spin structure is considered to be slightly disoriented. Therefore, the spins could be considered to have angles distributed about an average angle $\theta$. By choosing azimuthal angles as

$\theta_m = \theta + \varepsilon_m$ and $\theta_n = \theta + \varepsilon_n$

Where the $\varepsilon$'s are small positive or negative angular deviations.

Then, $\theta_m - \theta_n = \varepsilon_m - \varepsilon_n$ and $\theta_m + \theta_n = 2\theta + \varepsilon_m + \varepsilon_n$. After substituting these new angles in above equation number (1), the cosine and sine terms can be expanded up to the fourth order of $\varepsilon_m$ and $\varepsilon_n$ as following.

$$E(\theta) = E_0 + E(\varepsilon) + E(\varepsilon^2) + E(\varepsilon^3) + E(\varepsilon^4) + \ldots\ldots\ldots..$$

If the fifth and higher order perturbations are neglected, then

$$E(\theta) = E_0 + E(\varepsilon) + E(\varepsilon^2) + E(\varepsilon^3) + E(\varepsilon^4) \tag{2}$$

Here

$$E_0 = -\frac{1}{2}\sum_{m,n=1}^{N}\left(JZ_{|m-n|} - \frac{\omega}{4}\Phi_{|m-n|}\right) + \frac{3\omega}{8}\cos2\theta\sum_{m,n=1}^{N}\Phi_{|m-n|} - \cos^2\theta\sum_{m=1}^{N}D_m^{(2)}$$

$$-\cos^4\theta\sum_{m=1}^{N}D_m^{(4)} - N(H_{in}\sin\theta + H_{out}\cos\theta + K_s\sin2\theta) + \frac{N_dN^2}{\mu_0} \tag{3}$$

$$E(\varepsilon) = -\frac{3\omega}{8}\sin2\theta\sum_{m,n=1}^{N}\Phi_{|m-n|}(\varepsilon_m + \varepsilon_n) + \sin2\theta\sum_{m=1}^{N}D_m^{(2)}\varepsilon_m + 2\cos^2\theta\sin2\theta\sum_{m=1}^{N}D_m^{(4)}\varepsilon_m$$

$$-H_{in}\cos\theta\sum_{m=1}^{N}\varepsilon_m + H_{out}\sin\theta\sum_{m=1}^{N}\varepsilon_m - 2K_s\cos2\theta\sum_{m=1}^{N}\varepsilon_m \tag{4}$$

$$E(\varepsilon^2) = \frac{1}{4}\sum_{m,n=1}^{N}\left(JZ_{|m-n|} - \frac{\omega}{4}\Phi_{|m-n|}\right)(\varepsilon_m - \varepsilon_n)^2 - \frac{3\omega}{16}\cos2\theta\sum_{m,n=1}^{N}\Phi_{|m-n|}(\varepsilon_m + \varepsilon_n)^2$$

$$+\cos2\theta\sum_{m=1}^{N}D_m^{(2)}\varepsilon_m^2 + 2\cos^2\theta(\cos^2\theta - 3\sin^2\theta)\sum_{m=1}^{N}D_m^{(4)}\varepsilon_m^2 + \frac{H_{in}}{2}\sin\theta\sum_{m=1}^{N}\varepsilon_m^2$$

$$+\frac{H_{out}}{2}\cos\theta\sum_{m=1}^{N}\varepsilon_m^2 - \frac{N_d}{2\mu_0}\sum_{m,n=1}^{N}(\varepsilon_m - \varepsilon_n)^2 + 2K_s\sin2\theta\sum_{m=1}^{N}\varepsilon_m^2 \tag{5}$$

$$E(\varepsilon^3) = \frac{\omega}{16}\sin2\theta\sum_{m,n=1}^{N}\Phi_{|m-n|}(\varepsilon_m + \varepsilon_n)^3 - \frac{4}{3}\sin\theta\cos\theta\sum_{m=1}^{N}D_m^{(2)}\varepsilon_m^3$$

$$-4\sin\theta\cos\theta\left(\frac{5}{3}\cos^2\theta - \sin^2\theta\right)\sum_{m=1}^{N}D_m^{(4)}\varepsilon_m^3 + \frac{H_{in}}{6}\cos\theta\sum_{m=1}^{N}\varepsilon_m^3 - \frac{H_{out}}{6}\sin\theta\sum_{m=1}^{N}\varepsilon_m^3$$

$$+\frac{4}{3}K_s cos2\theta \sum_{m=1}^{N} \varepsilon_m^3 \tag{6}$$

$$E(\varepsilon^4) = -\frac{1}{48}\sum_{m,n=1}^{N}\left(JZ_{|m-n|} - \frac{\omega}{4}\Phi_{|m-n|}\right)(\varepsilon_m - \varepsilon_n)^4 + \frac{\omega}{64}cos2\theta \sum_{m,n=1}^{N}\Phi_{|m-n|}(\varepsilon_m + \varepsilon_n)^4$$

$$-\frac{1}{3}cos2\theta \sum_{m=1}^{N} D_m^{(2)}\varepsilon_m^4 - (\frac{5}{3}cos^4\theta - 8cos^2\theta sin^2\theta + sin^4\theta)\sum_{m=1}^{N} D_m^{(4)}\varepsilon_m^4$$

$$-\frac{H_{in}}{24}sin\theta \sum_{m=1}^{N}\varepsilon_m^4 - \frac{H_{out}}{24}cos\theta \sum_{m=1}^{N}\varepsilon_m^4 + \frac{N_d}{24\mu_0}\sum_{m,n=1}^{N}(\varepsilon_m - \varepsilon_n)^4$$

$$-\frac{2}{3}K_s sin2\theta \sum_{m=1}^{N}\varepsilon_m^4 \tag{7}$$

For films with three spin layers, $N = 3$. Therefore, $m$ and $n$ change from 1 to 3.

$$E_0 = -\frac{3}{2}\left(JZ_0 - \frac{\omega}{4}\Phi_0\right) - 2\left(JZ_1 - \frac{\omega}{4}\Phi_1\right) + \frac{3\omega}{8}cos2\theta(3\Phi_0 + 4\Phi_1) - cos^2\theta(D_1^{(2)} + D_2^{(2)}$$

$$+D_3^{(2)}) - cos^4\theta\left(D_1^{(4)} + D_2^{(4)} + D_3^{(4)}\right) - 3(H_{in}sin\theta + H_{out}cos\theta + K_s sin2\theta) + 9\frac{N_d}{\mu_0} \tag{8}$$

$$E(\varepsilon) = -\frac{3\omega}{4}sin2\theta[\Phi_0(\varepsilon_1 + \varepsilon_2 + \varepsilon_3) + \Phi_1(\varepsilon_1 + 2\varepsilon_2 + \varepsilon_3)] + sin2\theta(D_1^{(2)}\varepsilon_1 + D_2^{(2)}\varepsilon_2$$

$$+D_3^{(2)}\varepsilon_3) + 2cos^2\theta sin2\theta\left(D_1^{(4)}\varepsilon_1 + D_2^{(4)}\varepsilon_2 + D_3^{(4)}\varepsilon_3\right) - H_{in}cos\theta(\varepsilon_1 + \varepsilon_2 + \varepsilon_3)$$

$$+H_{out}sin\theta(\varepsilon_1 + \varepsilon_2 + \varepsilon_3) - 2K_s cos2\theta(\varepsilon_1 + \varepsilon_2 + \varepsilon_3) \tag{9}$$

$$E(\varepsilon^2) = \frac{1}{2}\left(JZ_1 - \frac{\omega}{4}\Phi_1\right)(\varepsilon_1^2 + 2\varepsilon_2^2 + \varepsilon_3^2 - 2\varepsilon_1\varepsilon_2 - 2\varepsilon_2\varepsilon_3) - \frac{3\omega}{8}cos2\theta[2\Phi_0(\varepsilon_1^2 + \varepsilon_2^2$$

$$+\varepsilon_3^2) + \Phi_1(\varepsilon_1^2 + 2\varepsilon_2^2 + \varepsilon_3^2 + 2\varepsilon_1\varepsilon_2 + 2\varepsilon_2\varepsilon_3)] + cos2\theta(D_1^{(2)}\varepsilon_1^2 + D_2^{(2)}\varepsilon_2^2$$

$$+D_3^{(2)}\varepsilon_3^2) + 2cos^2\theta(cos^2\theta - 3sin^2\theta)\left(D_1^{(4)}\varepsilon_1^2 + D_2^{(4)}\varepsilon_2^2 + D_3^{(4)}\varepsilon_3^2\right)$$

$$+\frac{H_{in}}{2}sin\theta(\varepsilon_1^2 + \varepsilon_2^2 + \varepsilon_3^2) + \frac{H_{out}}{2}cos\theta(\varepsilon_1^2 + \varepsilon_2^2 + \varepsilon_3^2) - 2\frac{N_d}{\mu_0}(\varepsilon_1^2 + \varepsilon_2^2$$

$$+\varepsilon_3^2 - \varepsilon_1\varepsilon_2 - \varepsilon_1\varepsilon_3 - \varepsilon_2\varepsilon_3) + 2K_s sin2\theta(\varepsilon_1^2 + \varepsilon_2^2 + \varepsilon_3^2) \tag{10}$$

$$E(\varepsilon^3) = \frac{\omega}{8}sin2\theta[4\Phi_0(\varepsilon_1^3 + \varepsilon_2^3 + \varepsilon_3^3) + \Phi_1(\varepsilon_1^3 + 3\varepsilon_1^2\varepsilon_2 + 3\varepsilon_1\varepsilon_2^2 + 2\varepsilon_2^3 + 3\varepsilon_2^2\varepsilon_3$$

$$+3\varepsilon_2\varepsilon_3^2 + \varepsilon_3^3)] - \frac{4}{3}sin\theta cos\theta\left(D_1^{(2)}\varepsilon_1^3 + D_2^{(2)}\varepsilon_2^3 + D_3^{(2)}\varepsilon_3^3\right) - 4sin\theta cos\theta$$

$$\left(\frac{5}{3}cos^2\theta - sin^2\theta\right)\left(D_1^{(4)}\varepsilon_1^3 + D_2^{(4)}\varepsilon_2^3 + D_3^{(4)}\varepsilon_3^3\right) + \frac{H_{in}}{6}cos\theta(\varepsilon_1^3 + \varepsilon_2^3 + \varepsilon_3^3)$$

$$-\frac{H_{out}}{6}sin\theta(\varepsilon_1^3 + \varepsilon_2^3 + \varepsilon_3^3) + \frac{4}{3}K_s cos2\theta(\varepsilon_1^3 + \varepsilon_2^3 + \varepsilon_3^3) \quad (11)$$

$$E(\varepsilon^4) = -\frac{1}{24}\left(JZ_1 - \frac{\omega}{4}\Phi_1\right)(\varepsilon_1^4 - 4\varepsilon_1^3\varepsilon_2 + 6\varepsilon_1^2\varepsilon_2^2 - 4\varepsilon_1\varepsilon_2^3 + 2\varepsilon_2^4 - 4\varepsilon_2^3\varepsilon_3 + 6\varepsilon_2^2\varepsilon_3^2$$

$$-4\varepsilon_2\varepsilon_3^3 + \varepsilon_3^4) + \frac{\omega}{32}cos2\theta[8\Phi_0(\varepsilon_1^4 + \varepsilon_2^4 + \varepsilon_3^4) + \Phi_1(\varepsilon_1^4 + 4\varepsilon_1^3\varepsilon_2 + 6\varepsilon_1^2\varepsilon_2^2$$

$$+4\varepsilon_1\varepsilon_2^3 + 2\varepsilon_2^4 + 4\varepsilon_2^3\varepsilon_3 + 6\varepsilon_2^2\varepsilon_3^2 + 4\varepsilon_2\varepsilon_3^3 + \varepsilon_3^4)] - \frac{1}{3}cos2\theta(D_1^{(2)}\varepsilon_1^4 + D_2^{(2)}\varepsilon_2^4$$

$$+D_3^{(2)}\varepsilon_3^4) - \left(\frac{5}{3}cos^4\theta - 8cos^2\theta sin^2\theta + sin^4\theta\right)\left(D_1^{(4)}\varepsilon_1^4 + D_2^{(4)}\varepsilon_2^4 + D_3^{(4)}\varepsilon_3^4\right)$$

$$-\frac{H_{in}}{24}sin\theta(\varepsilon_1^4 + \varepsilon_2^4 + \varepsilon_3^4) - \frac{H_{out}}{24}cos\theta(\varepsilon_1^4 + \varepsilon_2^4 + \varepsilon_3^4) + \frac{N_d}{6\mu_0}(\varepsilon_1^4 - 2\varepsilon_1^3\varepsilon_2$$

$$+3\varepsilon_1^2\varepsilon_2^2 - 2\varepsilon_1\varepsilon_2^3 + \varepsilon_2^4 - 2\varepsilon_1^3\varepsilon_3 + 3\varepsilon_1^2\varepsilon_3^2 - 2\varepsilon_1\varepsilon_3^3 + \varepsilon_3^4 - 2\varepsilon_2^3\varepsilon_3 + 3\varepsilon_2^2\varepsilon_3^2$$

$$-2\varepsilon_2\varepsilon_3^3) - \frac{2}{3}K_s sin2\theta(\varepsilon_1^4 + \varepsilon_2^4 + \varepsilon_3^4) \quad (12)$$

First order perturbation term can be expressed in terms of a row and column matrix with all seven terms in each as following.

$$E(\varepsilon) = \vec{\alpha}.\vec{\varepsilon}$$

Here terms of $\alpha$ are given by $\alpha_1, \alpha_2$ and $\alpha_3$.

$$\alpha_1 = -\frac{3\omega}{4}sin2\theta(\Phi_0 + \Phi_1) + sin2\theta D_1^{(2)} + 2cos^2\theta sin2\theta D_1^{(4)} - H_{in}cos\theta + H_{out}sin\theta - 2K_s cos2\theta \quad (13)$$

$$\alpha_2 = -\frac{3\omega}{4}sin2\theta(\Phi_0 + \Phi_1) + sin2\theta D_2^{(2)} + 2cos^2\theta sin2\theta D_2^{(4)} - H_{in}cos\theta + H_{out}sin\theta - 2K_s cos2\theta \quad (14)$$

$$\alpha_3 = -\frac{3\omega}{4}sin2\theta(\Phi_0 + \Phi_1) + sin2\theta D_3^{(2)} + 2cos^2\theta sin2\theta D_3^{(4)} - H_{in}cos\theta + H_{out}sin\theta - 2K_s cos2\theta \quad (15)$$

Second order perturbation term can be expressed in terms of a two by two matrix. A row matrix and a column matrix as following.

$$E(\varepsilon^2) = \frac{1}{2}\vec{\varepsilon}.C.\vec{\varepsilon}$$

Elements of $3 \times 3$ matrix ($C$) are delineated by

$$C_{11} = JZ_1 - \frac{\omega}{4}\Phi_1 - \frac{3\omega}{4}cos2\theta(2\Phi_0 + \Phi_1) + 2cos2\theta D_1^{(2)} + 4cos^2\theta(cos^2\theta - 3sin^2\theta)D_1^{(4)} + H_{in}sin\theta$$

$$+H_{out}cos\theta - \frac{4N_d}{\mu_0} + 4K_s sin2\theta \tag{16}$$

$$C_{12} = C_{21} = C_{23} = C_{32} = -JZ_1 + \frac{\omega}{4}\Phi_1 - \frac{3\omega}{4}cos2\theta\Phi_1 + \frac{2N_d}{\mu_0} \tag{17}$$

$$C_{13} = C_{31} = \frac{2N_d}{\mu_0} \tag{18}$$

$$C_{22} = 2(JZ_1 - \frac{\omega}{4}\Phi_1) - \frac{3\omega}{2}cos2\theta(\Phi_0 + \Phi_1) + 2cos2\theta D_2^{(2)} + 4cos^2\theta(cos^2\theta - 3sin^2\theta)D_2^{(4)}$$

$$+H_{in}sin\theta + H_{out}cos\theta - \frac{4N_d}{\mu_0} + 4K_s sin2\theta \tag{19}$$

$$C_{33} = JZ_1 - \frac{\omega}{4}\Phi_1 - \frac{3\omega}{4}cos2\theta(2\Phi_0 + \Phi_1) + 2cos2\theta D_3^{(2)} + 4cos^2\theta(cos^2\theta - 3sin^2\theta)D_3^{(4)}$$

$$+H_{in}sin\theta + H_{out}cos\theta - \frac{4N_d}{\mu_0} + 4K_s sin2\theta \tag{20}$$

Third order perturbation term can be expressed in terms of a two by two matrix. A row matrix and a column matrix as following.

$$E(\varepsilon^3) = \varepsilon^2.\beta.\vec{\varepsilon}$$

Elements of $3 \times 3$ matrix ($\beta$) are specified by

$$\beta_{11} = \frac{\omega}{8}sin2\theta(4\Phi_0 + \Phi_1) - \frac{4}{3}sin\theta cos\theta D_1^{(2)} - 4sin\theta cos\theta\left(\frac{5}{3}cos^2\theta - sin^2\theta\right)D_1^{(4)}$$

$$+\frac{H_{in}}{6}cos\theta - \frac{H_{out}}{6}sin\theta + \frac{4}{3}K_s cos2\theta \tag{21}$$

$$\beta_{12} = \beta_{21} = \beta_{23} = \beta_{32} = \frac{3\omega}{8}sin2\theta\Phi_1 \tag{22}$$

$$\beta_{13} = \beta_{31} = 0 \tag{23}$$

$$\beta_{22} = \frac{\omega}{4}sin2\theta(2\Phi_0 + \Phi_1) - \frac{4}{3}sin\theta cos\theta D_2^{(2)} - 4sin\theta cos\theta\left(\frac{5}{3}cos^2\theta - sin^2\theta\right)D_2^{(4)}$$

$$+\frac{H_{in}}{6}cos\theta - \frac{H_{out}}{6}sin\theta + \frac{4}{3}K_s cos2\theta \tag{24}$$

$$\beta_{33} = \frac{\omega}{8}sin2\theta(4\Phi_0 + \Phi_1) - \frac{4}{3}sin\theta cos\theta D_3^{(2)} - 4sin\theta cos\theta\left(\frac{5}{3}cos^2\theta - sin^2\theta\right)D_3^{(4)}$$

$$+\frac{H_{in}}{6}\cos\theta - \frac{H_{out}}{6}\sin\theta + \frac{4}{3}K_s\cos2\theta \tag{25}$$

Fourth order perturbation term can be expressed in terms of a two by two matrix. A row matrix and a column matrix as following.

$$E(\varepsilon^4) = \varepsilon^3.F.\vec{\varepsilon} + \varepsilon^2.G.\varepsilon^2$$

Elements of $3 \times 3$ matrix ($F$ and $G$) are delineated by

$$F_{11} = -\frac{1}{24}\left(JZ_1 - \frac{\omega}{4}\Phi_1\right) + \frac{\omega}{32}\cos2\theta(8\Phi_0 + \Phi_1) - \frac{1}{3}\cos2\theta D_1^{(2)}$$

$$-\left(\frac{5}{3}\cos^4\theta - 8\cos^2\theta\sin^2\theta + \sin^4\theta\right)D_1^{(4)} - \frac{H_{in}}{24}\sin\theta - \frac{H_{out}}{24}\cos\theta + \frac{N_d}{6\mu_0} - \frac{2}{3}K_s\sin2\theta \tag{26}$$

$$F_{12} = F_{21} = F_{23} = F_{32} = \frac{1}{6}\left(JZ_1 - \frac{\omega}{4}\Phi_1\right) + \frac{\omega}{8}\cos2\theta\Phi_1 - \frac{N_d}{3\mu_0} \tag{27}$$

$$F_{13} = F_{31} = -\frac{N_d}{3\mu_0} \tag{28}$$

$$F_{22} = -\frac{1}{12}\left(JZ_1 - \frac{\omega}{4}\Phi_1\right) + \frac{\omega}{16}\cos2\theta(4\Phi_0 + \Phi_1) - \frac{1}{3}\cos2\theta D_2^{(2)}$$

$$-\left(\frac{5}{3}\cos^4\theta - 8\cos^2\theta\sin^2\theta + \sin^4\theta\right)D_2^{(4)} - \frac{H_{in}}{24}\sin\theta - \frac{H_{out}}{24}\cos\theta + \frac{N_d}{6\mu_0} - \frac{2}{3}K_s\sin2\theta \tag{29}$$

$$F_{33} = -\frac{1}{24}\left(JZ_1 - \frac{\omega}{4}\Phi_1\right) + \frac{\omega}{32}\cos2\theta(8\Phi_0 + \Phi_1) - \frac{1}{3}\cos2\theta D_3^{(2)}$$

$$-\left(\frac{5}{3}\cos^4\theta - 8\cos^2\theta\sin^2\theta + \sin^4\theta\right)D_3^{(4)} - \frac{H_{in}}{24}\sin\theta - \frac{H_{out}}{24}\cos\theta + \frac{N_d}{6\mu_0} - \frac{2}{3}K_s\sin2\theta \tag{30}$$

$$G_{11} = G_{22} = G_{33} = 0 \tag{31}$$

$$G_{12} = G_{21} = -\frac{1}{8}\left(JZ_1 - \frac{\omega}{4}\Phi_1\right) + \frac{3\omega}{32}\cos2\theta\Phi_1 + \frac{N_d}{4\mu_0} \tag{32}$$

$$G_{23} = G_{32} = -\frac{1}{8}\left(JZ_1 - \frac{\omega}{4}\Phi_1\right) + \frac{3\omega}{32}\cos2\theta\Phi_1 + \frac{N_d}{2\mu_0} \tag{33}$$

$$G_{13} = G_{31} = \frac{N_d}{4\mu_0} \tag{34}$$

Therefore, the total magnetic energy given in equation (2) can be deduced to

$$E(\theta) = E_0 + \vec{\alpha}.\vec{\varepsilon} + \frac{1}{2}\vec{\varepsilon}.C.\vec{\varepsilon} + \varepsilon^2.\beta.\vec{\varepsilon} + \varepsilon^3.F.\vec{\varepsilon} + \varepsilon^2.G.\varepsilon^2 \tag{35}$$

For the minimum energy of the second order perturbed term

$$\vec{\varepsilon} = -C^+ . \alpha \tag{36}$$

Here $C^+$ is the pseudo inverse of matrix C. $C^+$ can be found using

$$C.C^+ = 1 - \frac{E}{N} \tag{37}$$

Here $E$ is the matrix with all elements given by $E_{mn} = 1$. $I$ is the identity matrix.

Therefore, from the matrix equation (36)

$$\varepsilon_1 = -(C_{11}^+ \alpha_1 + C_{12}^+ \alpha_2 + C_{13}^+ \alpha_3) \tag{38}$$

$$\varepsilon_2 = -(C_{21}^+ \alpha_1 + C_{22}^+ \alpha_2 + C_{23}^+ \alpha_3) \tag{39}$$

$$\varepsilon_3 = -(C_{31}^+ \alpha_1 + C_{32}^+ \alpha_2 + C_{33}^+ \alpha_3) \tag{40}$$

After substituting ε in equation (35), the total magnetic energy can be determined.

## 3. Results and Discussion:

All the graphs in this manuscript were plotted for ferromagnetic films with body centered cubic lattice and three spin layers. For ferromagnetic films with bcc (001) structure, $Z_0=0$, $Z_1=4$, $Z_2=0$, $\Phi_0 = 5.8675$ and $\Phi_1 = 2.7126$ [14-16]. 3D plot of energy versus angle and stress induced anisotropy constant is given in figure 1 for $\frac{D_1^{(2)}}{\omega} = 5$, $\frac{D_2^{(2)}}{\omega} = 10$ and $\frac{D_3^{(2)}}{\omega} = 10$. Here other parameters are fixed at $\frac{J}{\omega} = \frac{D_1^{(4)}}{\omega} = \frac{D_2^{(4)}}{\omega} = \frac{D_3^{(4)}}{\omega} = \frac{H_{in}}{\omega} = \frac{N_d}{\mu_0 \omega} = \frac{H_{out}}{\omega} = 10$. for this simulation. The energy maximums can be observed at $\frac{K_s}{\omega} = 2, 23, 48, 71, 94$. The major maximum observed at about $\frac{K_s}{\omega} = 71$. Energy in these graphs is in the order of $10^{19}$. The maximums energy values observed in this graph are approximately the same when compared to the sc structured ferromagnetic thin films with the same spin layers [17]. The peaks along the axis of angle are closely packed in the fourth order perturbed case compared to the second and third order perturbed cases [6, 10, 13]. The shape of the graph is also entirely different from the graphs obtained using the second and third order perturbed Heisenberg Hamiltonian [6, 10, 14]. The peaks are periodically distributed. The total magnetic energy slightly changes when compared to sc structured ferromagnetic thin films with the same spin layers [17]. But it was found ($10^{51}$) significantly higher in the two spin layer for the same structure [18].

Figure 2 shows the graph of energy versus angle for $\frac{K_S}{\omega} = 71$ after fixing the other parameters at the values given above. In this graph, magnetic easy directions can be observed at 0.754 and 3.864 radians. The energy maximums can be found at 2.356 and 5.404 radians. The angle between consecutive magnetic easy and hard directions is approximately 90 degrees.

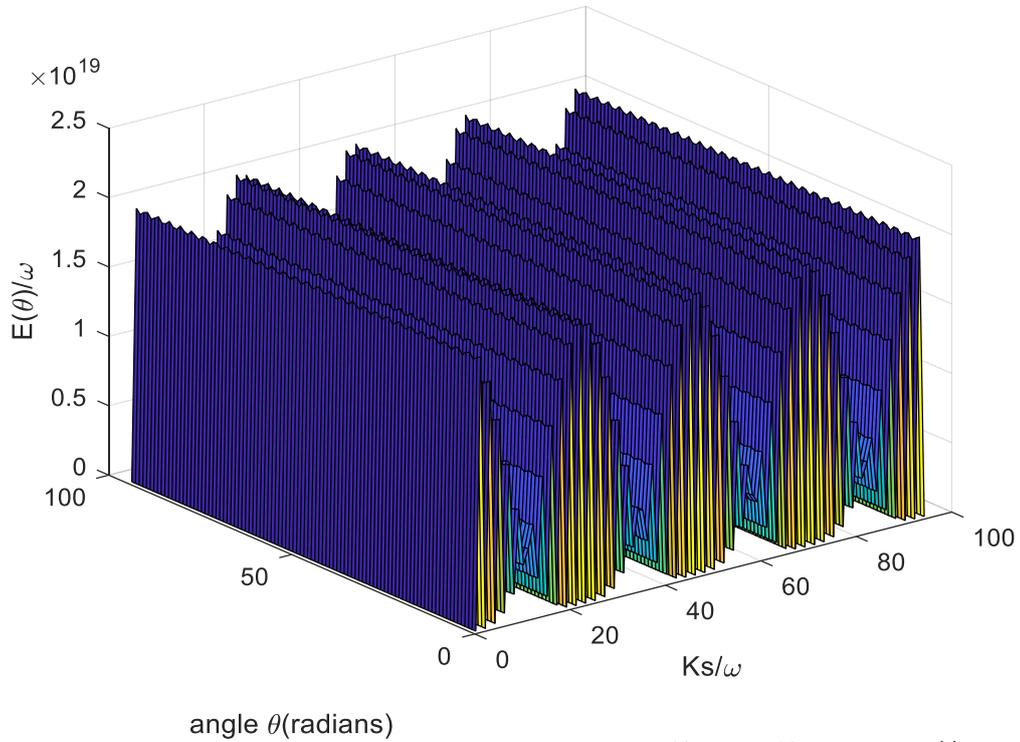

**Figure 1:** 3D plot of energy versus angle and stress induced anisotropy for $\frac{D_1^{(2)}}{\omega} = 5$, $\frac{D_2^{(2)}}{\omega} = 10$ and $\frac{D_3^{(2)}}{\omega} = 10$.

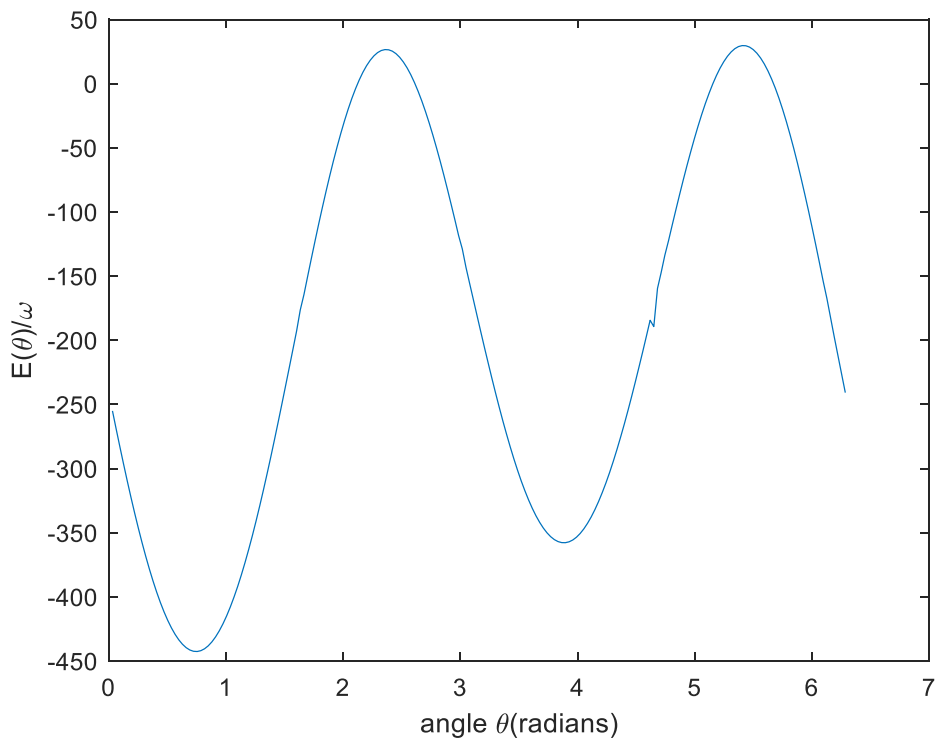

**Figure 2:** 2D graph of energy versus angle for $\frac{K_S}{\omega} = 71$.

Figure 3 represents the 3D plot of energy versus angle and stress induced anisotropy constant for $\frac{D_1^{(2)}}{\omega} = 10$, $\frac{D_2^{(2)}}{\omega} = 5$ and $\frac{D_3^{(2)}}{\omega} = 10$. Here other parameters are fixed at $\frac{J}{\omega} = \frac{D_1^{(4)}}{\omega} = \frac{D_2^{(4)}}{\omega} = \frac{D_3^{(4)}}{\omega} = \frac{H_{in}}{\omega} = \frac{N_d}{\mu_0 \omega} = \frac{H_{out}}{\omega} = 10$ for this simulation. In this graph, the energy maximums can be observed at $\frac{K_s}{\omega} = 2, 23, 48, 71$ and $94$. The major maximum was observed at about $\frac{K_s}{\omega} = 71$. Energy in these graphs is in the order of $10^{11}$. Because there are several energy maxima and minima in this 3D plot, it implies that the films with some stress induced anisotropy can be easily magnetized. The same maximums energy values are observed in this graph when the bottom and middle spin layers are interchanged. Additionally, the total magnetic energy also decreases. The shape of the 3D plots obtained for the ferromagnetic film with two spin layers using the fourth order perturbed Heisenberg Hamiltonian with all seven magnetic energy parameters is different from these 3D plots [18].

Figure 4 shows the graph of energy versus angle for $\frac{K_s}{\omega} = 71$. In this graph, energy minimums can be observed at 0.754 and 3.896 radians. The major minimum was observed at about 0.754 radians. The energy maximums can be found at 2.356 and 5.404 radians. The major maximum was observed at about 5.404 radians. Therefore, the angle between consecutive magnetic easy and hard directions is nearly 90 degrees. For the ferromagnetic films with two spin layers, the angle between the magnetic easy and hard directions was exactly 90 degrees [18].

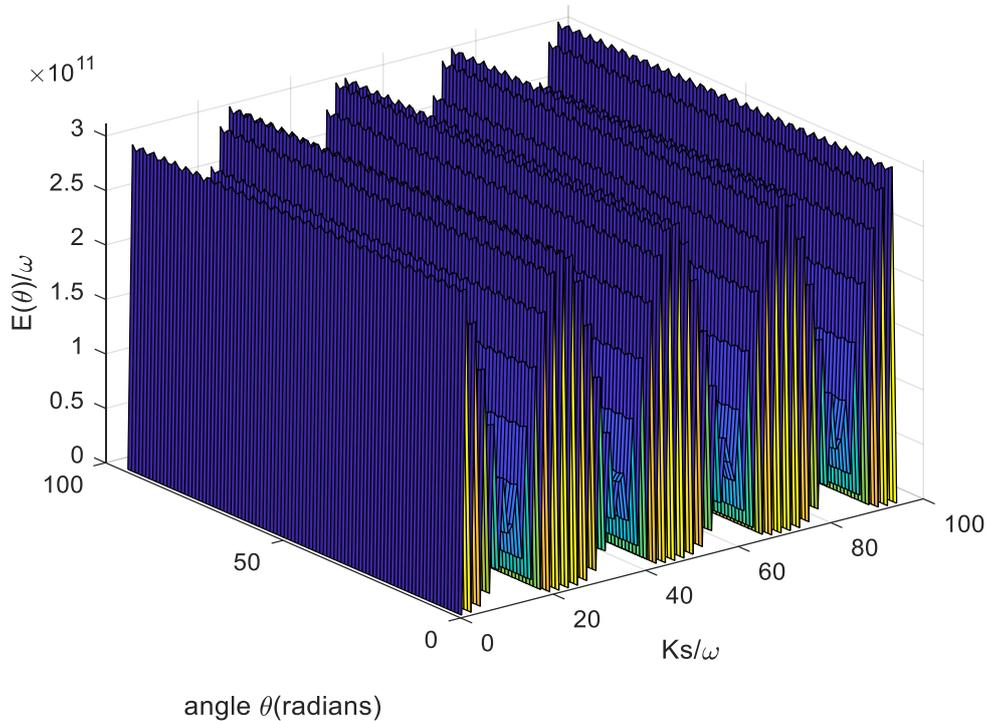

**Figure 3:** 3D plot of energy versus angle and stress induced anisotropy for $\frac{D_1^{(2)}}{\omega} = 10$, $\frac{D_2^{(2)}}{\omega} = 5$ and $\frac{D_3^{(2)}}{\omega} = 10$.

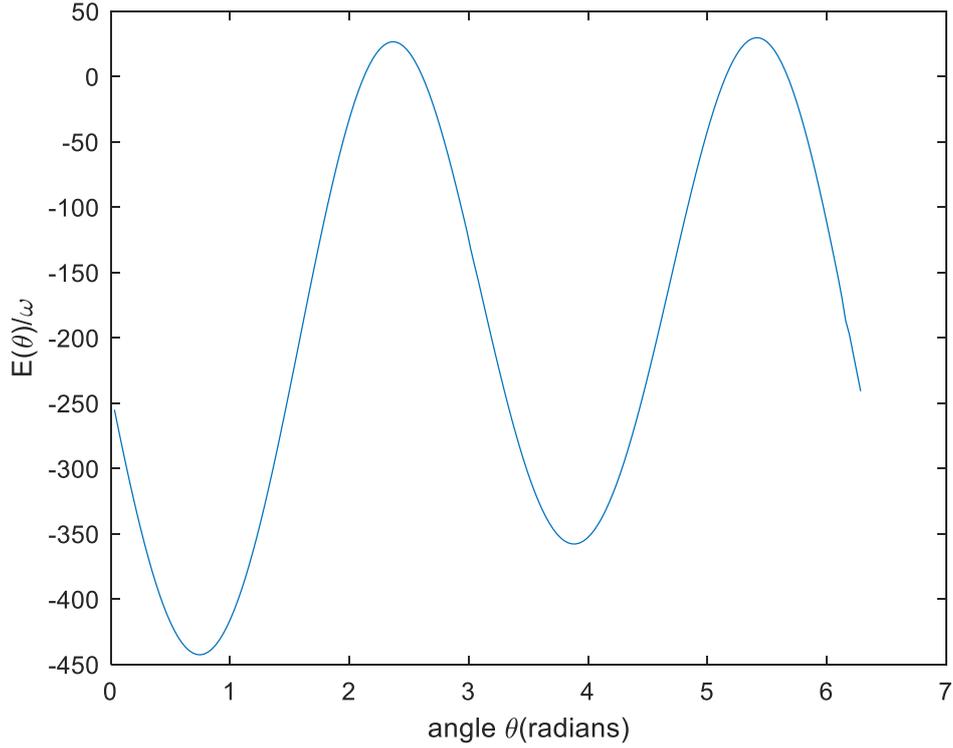

**Figure 4:** 2D graph of energy versus angle for $\frac{K_s}{\omega} = 71$.

Figure 5 represents the 3D plot of energy versus angle and stress induced anisotropy constant for $\frac{D_1^{(2)}}{\omega} = 10$, $\frac{D_2^{(2)}}{\omega} = 10$ and $\frac{D_3^{(2)}}{\omega} = 5$. Here other parameters are fixed at $\frac{J}{\omega} = \frac{D_1^{(4)}}{\omega} = \frac{D_2^{(4)}}{\omega} = \frac{D_3^{(4)}}{\omega} = \frac{H_{in}}{\omega} = \frac{N_d}{\mu_0 \omega} = \frac{H_{out}}{\omega} = 10$ for this simulation. In this graph, the energy maximums can be observed at $\frac{K_s}{\omega} = 2$, 23, 48, 71 and 94. The major maximum was observed at about $\frac{K_s}{\omega} = 71$. The total energy is in the order of $10^{21}$. According to Figures 1, 2 and 3, when the second order anisotropy constant in the top spin layer is less than those of the bottom and middle spin layers, the total magnetic energy increases. Magnetic materials with higher and lower magnetic energies can be found in applications of permanent and soft magnets, respectively. The films with lower magnetic anisotropy at the middle spin layer can be used as soft magnets, while the films with lower magnetic anisotropy at the top spin layer can be used as permanent magnets. The values of stress induced anisotropy at the maxima of 3D plots are exactly the same, when the values of the second order magnetic anisotropy constant of the top and bottom spin layers are interchanged. Some experimental studies reveal that the magnetic anisotropy constants vigorously govern the orientation of the magnetic easy axis of magnetic thin films. The minimum value of the energy is zero in all the 3D plots.

Figure 6 shows the graph of energy versus angle for $\frac{K_S}{\omega} = 71$. In this graph, energy minimums can be observed at 0.754 and 3.864 radians. The major minimum was observed at about 0.754 radians. The energy maximums can be found at 2.356 and 5.404 radians. The major maximum was observed at about 5.404 radians. Therefore, the angle between consecutive magnetic easy and hard directions is nearly 90 degrees. According to Figures 2, 4 and 6, the magnetic easy and hard directions were observed to have exactly the same values, when the second order magnetic anisotropy constant of the spin layers' changes.

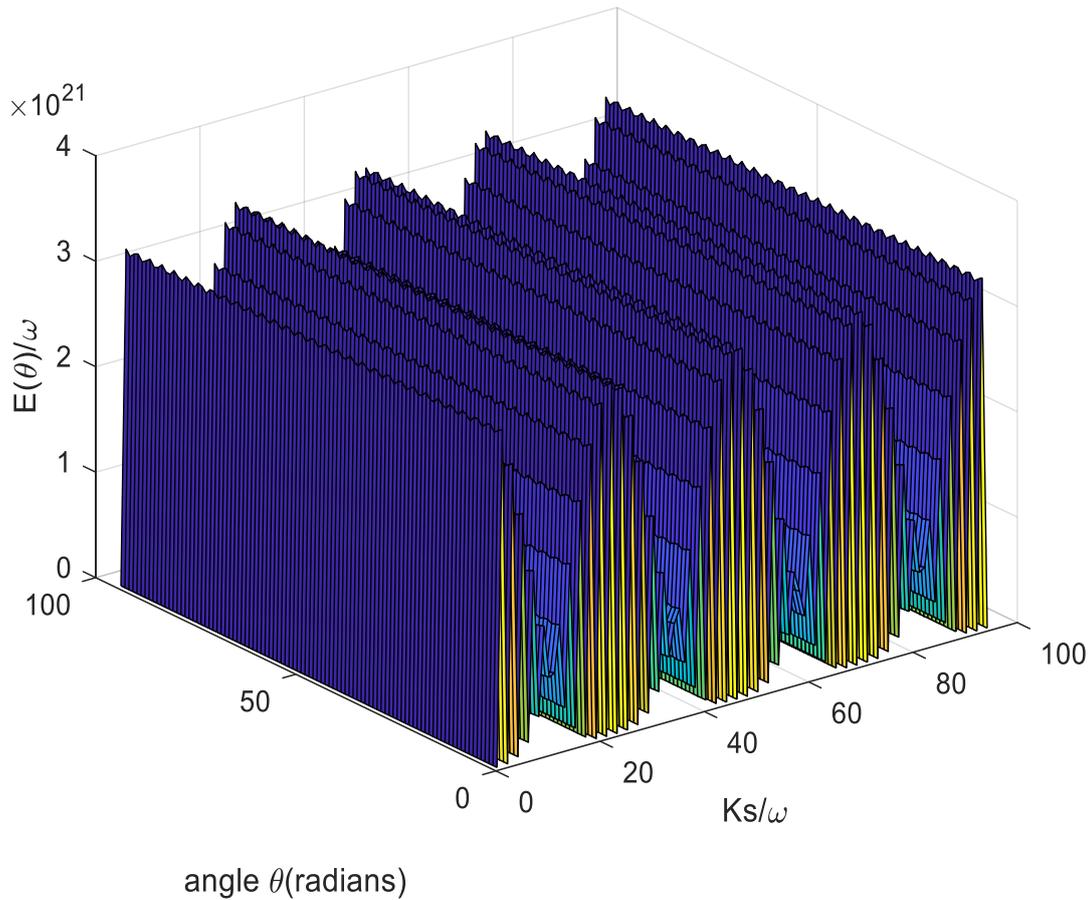

**Figure 5:** 3D plot of energy versus angle and stress induced anisotropy for $\frac{D_1^{(2)}}{\omega} = 10$, $\frac{D_2^{(2)}}{\omega} = 10$ and $\frac{D_3^{(2)}}{\omega} = 5$

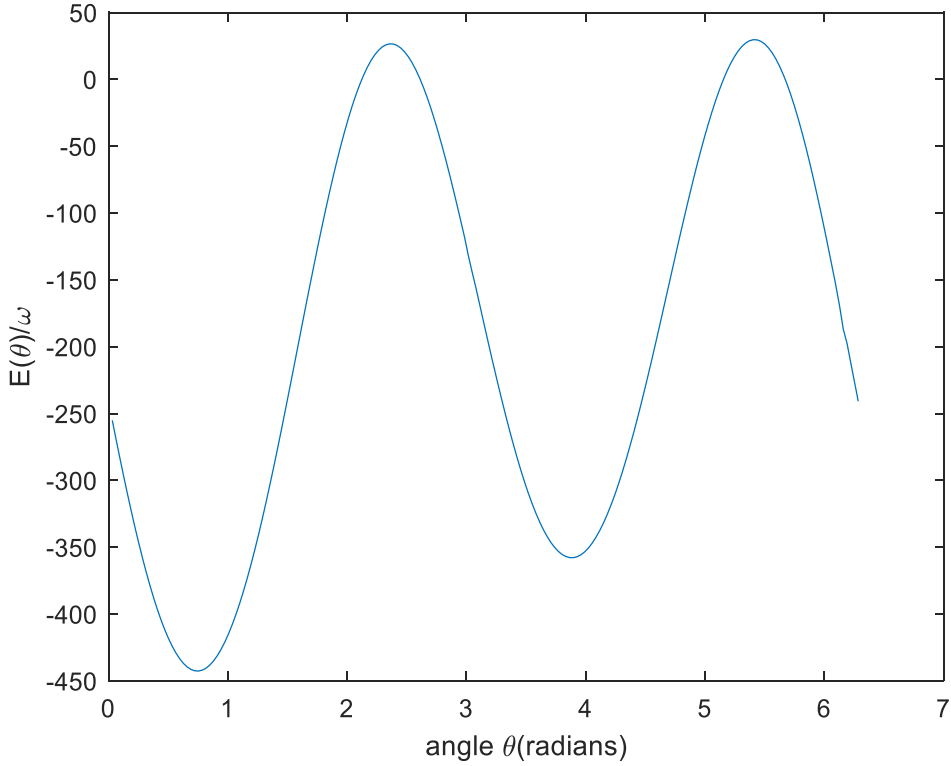

**Figure 6:** 2D graph of energy versus angle for $\frac{K_S}{\omega} = 71$.

## 4. Conclusion:

The magnetic properties of fourth order perturbed Heisenberg Hamiltonian equation with all seven magnetic energy parameters were discussed for body centered cubic structured ferromagnetic thin films with three spin layers. All the graphs between total magnetic energy versus angle and stress induced anisotropy constant were plotted using the fourth order perturbed Heisenberg Hamiltonian. According to 3D plots, the order of the magnetic energy is in the range from $10^{11}$ to $10^{21}$. The minimum order of magnetic energy ($10^{11}$) was observed, when the second order magnetic anisotropy constant in the middle spin layer is less than those of the bottom and top spin layers. Here the energy maximums were observed at $\frac{K_S}{\omega} =$ 2, 23, 48, 71 and 94. For $\frac{K_S}{\omega} = 71$, energy minimums were observed at 0.754 and 3.896 radians and the energy maximums were found at 2.356 and 5.404 radians. The maximum order of magnetic energy ($10^{21}$) was observed, when the second order magnetic anisotropy constant in the top spin layer is less than those of the bottom and middle spin layers. Here the energy maximums were observed at $\frac{K_S}{\omega} =$ 2, 23, 48, 71 and 94. For $\frac{K_S}{\omega} = 71$, energy minimums were observed at 0.754 and 3.864 radians, and the energy maximums were found at 2.356 and 5.404 radians. In all cases, the values of stress-induced anisotropy at the maxima of 3D plots are exactly the same when the values of the second-order magnetic anisotropy constants of the bottom, middle, and top spin layers are interchanged with each other. In 2D plots, the angle between consecutive magnetic easy and hard directions is approximately 90 degrees in all cases. Additionally, the magnetic easy and hard directions were observed to have exactly the same values when the second-order magnetic anisotropy constant of the spin layers' changes.

**References**

1. Zhao, D., Feng Liu, Huber, D.L., Lagally, M.G., Step-induced magnetic-hysteresis anisotropy in ferromagnetic thin films. *Journal of Applied Physics* **2002**, 91(5), 3150.

2. Bentaleb, M., Aouad El, N., Saber, M., Magnetic properties of the spin -1/2 Ising Ferromagnetic thin films with alternating superlattice configuration. *Chinese Journal of Physics* **2002**, 40(3), 307.

3. Spisak, D., Hafner, J., Theoretical study of FeCo/W (110) surface alloys. *Journal of Magnetism and Magnetic Materials* **2005**, 286, 386.

4. Ernst, A., Lueders, M., Temmerman, W.M., Szotek, Z., Van der Laan, G., Theoretical study of magnetic layers of nickel on copper; dead or live? *Journal of Physics: Condensed matter* **2000**, 12(26), 5599.

5. Samarasekara, P., Cadieu, F.J., Magnetic and Structural Properties of RF Sputtered Polycrystalline Lithium Mixed Ferrimagnetic Films. *Chinese journal of Physics* **2001**, 39(6), 635-40.

6. Samarasekara, P., De Silva, S.N.P., Heisenberg Hamiltonian solution of thick ferromagnetic films with second order perturbation. *Chinese Journal of Physics* **2007**, 45(2-I), 142.

7. Samarasekara P. and Saparamadu Udara, In plane oriented Strontium ferrite thin films described by spin reorientation. *Research & Reviews: Journal of Physics-STM journals* **2013**, 2(2), 12.

8. Samarasekara, P., Ekanayake, E.M.P., Ferromagnetic thin films with simple cubic structure as described by fourth order perturbed Heisenberg Hamiltonian. *STM journals: Journal of Nanoscience, NanoEngineering & Applications* **2020**, 10(3), 35-43.

9. Samarasekara, P., Mendoza, W.A., Effect of third order perturbation on Heisenberg Hamiltonian for non-oriented ultra-thin ferromagnetic films. *Electronic Journal of Theoretical Physics* **2010**, 7(24), 197.

10. Samarasekara, P., Warnakulasooriya, B.I., Five layered fcc ferromagnetic films as described by modified second order perturbed Heisenberg Hamiltonian. *Journal of science: University of Kelaniya* **2016**, 11, 11-21.

11. Yapa, N.U.S., Samarasekara, P., Sunil Dehipawala, Variation of magnetic easy direction of bcc structured ferromagnetic films up to thirty spin layers. *Ceylon Journal of Science* **2017**, 46(3), 93.

12. Samarasekara, P., Abeyratne, M.K., Dehipawalage, S., Heisenberg Hamiltonian with Second Order Perturbation for Spinel Ferrite Thin Films. *Electronic Journal of Theoretical Physics* **2009**, 6(20), 345.

13. Samarasekara, P., Influence of third order perturbation on Heisenberg Hamiltonian of thick ferromagnetic films. *Electronic Journal of Theoretical Physics* **2008**, 5(17), 227

14. Hucht, A., Usadel, K.D., Reorientation transition ultrathin ferromagnetic films. *Physical Review B* **1997**, 55, 12309.

15. Hucht, A., Usadel, K.D., Theory of the spin reorientation transition of ultra-thin ferromagnetic films. *Journal of Magnetism and Magnetic Materials* **1999**, 203(1), 88.

16. Usadel, K.D., Hucht, A., Anisotropy of ultrathin ferromagnetic films and the spin reorientation transition. *Physical Review B* **2002**, 66, 024419.

17. Farhan, M.S.M., Samarasekara, P., Investigation of magnetic properties of ferromagnetic thin films with three spin layers using fourth order perturbed Heisenberg Hamiltonian. *Ceylon Journal of Science* **2023**, 52(2), 133-142.

18. Farhan, M.S.M., Samarasekara, P., Fourth Order Perturbed Heisenberg Hamiltonian with Seven Magnetic Parameters of bcc Structured Ferromagnetic Ultrathin Films. *GESJ: Physics* **2022**, 1(26), 29-45.